%% file: secretary.tex
\documentclass[a4paper,UKenglish]{lipics}
\usepackage{color}
\usepackage[boxed]{algorithm2e}
\usepackage{mathrsfs}

\usepackage{amssymb}
\usepackage{amsmath}

\usepackage{verbatim}
 \newtheorem{proposition}[theorem]{Proposition}

\DeclareMathOperator*{\argmax}{\arg\!\max}

\def \xi {\mathbf{x}_{-i}}
\def \opt {\mathrm{OPT}}
\def \alg {\mathrm{ALG}}
\def \sim {\mathrm{SIMULATE}}
\def \online {\mathrm{ONLINE}}
\def \greedy {\mathrm{GREEDY}}

\def \cA {\mathcal{A}}

\def \cR {\mathcal{R}}

\def \real {\mathcal{R}}
\def \one {\mathbf{1}}

        {\hspace*{\fill}$\Box$\par}

\def \xi {\mathbf{x}_{-i}}
\def \opt {\mathrm{OPT}}
\def \alg {\mathrm{ALG}}

\def \cP {\mathcal{P}}

\def \cI {\mathcal{I}}
\def \cM {\mathcal{M}}
\def \E  {\mathbb{E}}
\def \pr {\mathrm{Pr}}
\def \cF {\mathcal{F}}

\title{The Simulated Greedy Algorithm for Several Submodular Matroid Secretary
  Problems}
\titlerunning{Algorithms for Submodular Matroid Secretary Problems}
\author[1]{Tengyu Ma}
\author[2]{Bo Tang}
\author[3]{Yajun Wang}
\affil[1]{Princeton University$^*$\\
 \texttt{tengyu@cs.princeton.edu}}
\affil[2]{University of Liverpool\footnote{This work was done when the authors were visiting
  Microsoft Research Asia.}\\
  \texttt{tangbonk1@gmail.com}}
\affil[3]{Microsoft Research Asia\\
  \texttt{yajunw@microsoft.com}}

\Copyright[nc-nd]
          {Bo Tang and Yajun Wang}

\begin{document}
\maketitle
\begin{abstract}
We study the matroid secretary problems with submodular valuation
functions. In these problems, the elements arrive in random
order. When one element arrives, we have to make an immediate and irrevocable decision on 
whether to accept it or not. The set of accepted elements must form an {\em independent set} in a predefined
matroid.
Our objective is to maximize the value of the accepted elements. In this paper, we focus on the case that the
valuation function is a non-negative and monotonically non-decreasing
submodular function. 

We introduce a general algorithm for such {\em submodular matroid
secretary problems}. In particular, we obtain constant competitive 
algorithms for the cases of laminar matroids and transversal
matroids. Our algorithms can be further applied to any independent set system defined
by the intersection of a {\em constant} number of laminar matroids, while
still achieving constant competitive ratios. Notice that laminar matroids
generalize uniform matroids and partition matroids.

On the other hand, when the underlying valuation function is linear, our
algorithm achieves a competitive ratio of $9.6$ for laminar matroids, 
which significantly improves the previous
result.

%
\end{abstract}
\setcounter{page}{0}
\thispagestyle{empty}
\newpage

\section{Introduction}

In the classical
secretary problem~\cite{Dynkin63,Freeman1983,Gardner60}, one
interviewer is interviewing $n$ candidates for a secretary
position. The candidates arrive in an online fashion and the
interviewer has to decide whether or not to hire the current candidate
when he/she arrives. The goal is to hire the best secretary. It has been shown that when the candidates are arriving in random order, there exists an algorithm that hires the best candidate with probability $1/e$, where $e$ is the base of the natural logarithm.

Recently, Babaioff et al.~\cite{BabaioffIK07} formulated the matroid
secretary problem. Instead of hiring one candidate (element), in the
matroid secretary problem, we seek to select a set of elements which form
an independent set in a matroid. Again, the elements arrive in
random order and the weights of the elements are revealed when they
arrive. When one element arrives, we have to make an immediate and irrevocable decision
on whether to accept this element or not. The important constraint is that the set of
accepted elements must form an independent set in the predefined matroid.
The objective is to maximize the total weights of the
selected elements. Notice that the decision on accepting a particular element will impact
our ability in accepting future elements.

In the matroid secretary problem, the value of a set of elements is the summation of the weights 
on these elements, i.e., the valuation function is linear.
In some
applications, however, it is more natural to measure the quality of a set by a
valuation function, which is not necessarily linear. One
set of functions widely used in the optimization community are the {\em
  submodular} functions.
Such functions are characterized as functions with diminishing
returns. We give the formal definition in Section \ref{sec:pre}.

For example, consider the following scenario. An advertiser is
targeting a few platforms to reach a good coverage of
audience. However, the coverage from different platforms may overlap
with each other. In this case, the performance of a particular set of
platforms can only be modelled as a submodular function. Assume the
advertiser has to negotiate with the platforms one by one in an
online fashion and has a
hard budget limit on targeting at most $k$ platforms. This is exactly
the matroid secretary problem with a submodular valuation function on a uniform
matroid. 

We can also consider multiple arriving
advertisers, while assuming platforms are available offline. One can
impose constraints both on the advertisers and platforms, e.g., each advertiser can
afford $k$ platforms, and each platform can support at most $\ell$
advertisers. This scenario can be modelled as an intersection of two
partition matroids, with a submodular valuation function, where the
objective is to maximize the value of an overall online assignment.

In this paper, we extend the matroid secretary problem to
the case with submodular valuation functions. In other words, the weights are
not directly associated with elements. Instead, there exists an
oracle
to query the value of any subset of the elements we have
seen. Our objective is to accept a set of elements which are independent in a given matroid with maximum
value with respect to a submodular valuation function. We refer such problems as {\em submodular matroid secretary problems}. We refer the original matroid secretary problems, i.e., those with linear valuation functions, as {\em linear matroid secretary problems}.

We use the competitive analysis to measure the performance of our algorithms following the matroid secretary problem
literature.
More formally, let $U$ be the set of elements and ${\mathcal M}$ be a matroid defined on $U$.
Before the process starts, an adversary assigns a submodular valuation function $f(\cdot): 2^{|U|} \rightarrow \real^+\cup \{0\}$, which maps any subset of $U$ to a non-negative real number. 
After that, there is a random permutation applied
to the elements to decide their arriving order to our online
algorithm. Our algorithm can only query $f(\cdot)$ using elements that have been seen. In other words,
the algorithm does not know $f(\cdot)$ before any element arrives.


Let $\opt_f(\cM) = \max_{S\in \cM} f(S)$ be the value of the optimal
independent set. The
objective of the submodular matroid secretary problem is to find an algorithm $\mathrm{Alg}$
which maximizes the following ratio: 
\begin{equation}
\label{eqn:competitive}
\inf_{f} \frac{ 
\E_{\cP, \cA}[f(\mathrm{Alg}_f(\cP, \cA)) ]
}{\opt_f(\cM)},
\end{equation}
where $\mathrm{Alg}_f(\cP,\cA)$ is the solution generated by
the algorithm given permutation $\cP$ and the internal randomness
$\cA$ of the algorithm with valuation function $f(\cdot)$. The expectation is taken over all
permutations and the internal randomness of the algorithm. We call the algorithm
is $C$-competitive, i.e., with competitive ratio $C$, if the ratio in
Eqn.(\ref{eqn:competitive}) is at least $1/C$.

{\bf Our contributions.} In this paper, we study the submodular
matroid secretary problem with submodular valuation functions that are {\em non-negative} and {\em monotonically
  non-decreasing}. Our contribution is two-fold.
First, we develop a general {\em simulated
  greedy} algorithm, which is inspired by the algorithm for the
linear matroid
secretary problem with transversal matroids
in~\cite{DimitrovP08,KorulaP09}. Our algorithm is constant
competitive for the submodular matroid secretary problem
with laminar matroids and transversal matroids. Our analysis can be
extended to the case that the independent set is defined as the intersection of a {\em constant} number of laminar
matroids. Notice that laminar matroids generalize uniform matroids and
partition matroids. When applying to the linear matroid secretary problem on
laminar matroids, our algorithm improves the competitive ratio from $\frac{16000}{3}$~\cite{ImW11} to $9.6$.
Our algorithm is also much simpler than the one in~\cite{ImW11}.

Second, our technique in analyzing submodular functions could be of
independent interest. Consider our simulated greedy algorithm for the
uniform matroid case with cardinality $\mu$. We maintain two sets $M$
and $N$, which are initially empty. In each time, we will select an
element $e\in U\setminus (M\cup N)$ such that $f_{M}(e)$ is maximized until $|M|=\mu$, where
$f(\cdot)$ is the
valuation function. With probability $p$, $e$ is placed into
$M$. Otherwise, i.e., with probability $1-p$, $e$ is placed into
$N$. We develop machineries to show that $\E[f(N)] = \Theta(\E[f(M)])$,
despite the fact that the elements are greedily selected with optimal marginal values against $M$. This fact is not intuitive though very important in
our analysis. See our result in
Section~\ref{sec:laminar} for more details.

{\bf Related work.}
The secretary problem has been studied decades ago. It is first
published in~\cite{Gardner60} and has been folklore even
earlier~\cite{Ferguson1989}. 
Several results have appeared to generalize the classical secretary
problem, while assuming that the elements arrive in random order. 
For example, Kleinberg~\cite{Kleinberg05} gave a
$1+O(1/\sqrt k)$-competitive algorithm for selecting at most $k$
elements to maximize the sum of the weights. Babaioff et
al.~\cite{BabaioffIKK07} provided a constant competitive algorithm for the Knapsack secretary problem, in
which each element has a weight and a size, and the objective is to accept a
set of elements whose total size is at most a given integer such that
the total weight is maximized. 


Babaioff et al.~\cite{BabaioffIK07} systematically introduced the {\em
matroid secretary problem}. 
The objective is to maximize the total weight of the selected
elements $S$, which form an independent set in a given matroid. They gave
an $O(\log r)$-competitive algorithm for a general 
matroid, i.e., the expected total weight of the elements in $S$ is $O (1/\log
r)$ of the optimal solution, where $r$ is the rank of the matroid. The competitive 
ratio has been recently improved to $O(\sqrt{\log{r}})$ by Chakraborty et
al.~\cite{ChakrabortyL2012}. However, the conjecture that the matroid
secretary problem with a general matroid allows a constant competitive algorithm is still
widely open, while constant competitive algorithms have been found for
various matroids: uniform/partition matroids
\cite{BabaioffIKK07,Kleinberg05}, truncated partition matroids
\cite{BabaioffIK07}, graphical matroids
\cite{BabaioffDGIT09,KorulaP09}, transversal matroids
\cite{DimitrovP08,KorulaP09}, laminar matroids~\cite{ImW11}, and regular and decomposable  matroids~\cite{DBLP:journals/corr/abs-1207-5146}.
%
%
For general matroids,
Soto~\cite{Soto11} developed a constant-competitive algorithm in
{\em random assignment model}, i.e., the weights of the elements are
assigned uniformly at random. This result can be extended to the
case where the elements arrive in an adversarial order \cite{GharanV11}.

Gupta et
al.~\cite{GuptaRST10} studied the {\em non-monotone} submodular
matroid maximization problem for both offline and online (secretary)
versions. For the online (secretary) version, they provided a $O(\log
r)$-competitive algorithm for general matroids and a constant
competitive algorithm for uniform matroids (algorithms achieving
constant competitive ratios are obtained independently by Bateni et
al.\cite{BateniHZ10}) and partition matroids.
Feldman et al.~\cite{Feldman2011}
developed a simpler algorithm with a better competitive ratio for partition matroids
for {\em monotonically non-decreasing} submodular functions.

{\bf Structure.} In Section~\ref{sec:pre}, we present some
preliminaries and our algorithm. We then analyze a simple stochastic
process in Section~\ref{sec:uniform}, which serves as a building
block for later analysis. In Section~\ref{sec:laminar}, we analyze the
algorithm for the cases of laminar matroids and the intersection of
constant number of laminar matroids. We discuss the transversal matroid case in Section~\ref{sec:transversal}.
 We
conclude with Section~\ref{sec:conclusion}. 

\section{Preliminaries}
\label{sec:pre}

\subsection{Matroids}
\def \mm {\mathcal{M}}
\def \mi {\mathcal{I}}

In the matroid secretary problem, the set of accepted elements must
form an independent set defined by a given matroid.

\begin{definition}[Matroids]
Let $U \neq \emptyset$ be the ground set and $\mi$ be a set of subsets
of $U$. The system $\mm = (U, \mi)$ is a matroid with independent
sets $\mi$ if:
\begin{enumerate}
\item If $A\subseteq B\subseteq U$ and $B\in \mi$, then $A\in \mi$.
\item If $A, B\in \mi$ and $|A| < |B|$, there exists an element $x \in
 B\setminus A$ such that $A\cup \{x\} \in \mi$.
\end{enumerate}
\end{definition}

In this paper, we work with the following two matroids.

\begin{definition}[Laminar matroids]
Let $U \neq \emptyset$ be the ground set. Let $\cF = \{B_1,\ldots,
B_\ell\}$ be a family of subsets
over $U$. $\cF$ is a laminar family, if for any $B_i, B_j$ such that
$|B_i|\leq |B_j|$, either $B_i\cap B_j =\emptyset$ or $B_i\subseteq
B_j$. Each set $B_i\in \cF$ is associated with capacity
$\mu(B_i)$. The laminar family $\cF$ and $\mu(\cdot)$ define a matroid $\mm = (U,
\mi)$, such that any set $T\subseteq U$ is independent if for all
$1\leq i\leq \ell$, $|T\cap B_i| \leq \mu(B_i)$.
\end{definition}

In particular, each $B_i$ defines a capacity constraint on the
independent sets and a set is independent if it satisfies all such
constraints. 
For simplicity, we assume all $B_i$s are
distinct and $\mu(B_i) < \mu(B_j)$ if $B_i\subset B_j$. Otherwise, the
capacity constraint in $B_i$ is redundant.

\begin{definition}[Transversal matroids]
Let $G=(L,R,E)$ be an undirected bipartite graph with left nodes 
$L$, right nodes $R$ and edges $E$. In the transversal matroid
defined by $G$, the ground set is $L$ and a set of left nodes
$S\subseteq L$ is independent if there exists a matching in $G$ such
that the set of left nodes in the matching is $S$.
\end{definition}


\subsection{Submodular functions}

In this paper, we assume the quality of the solution is measured by a
submodular function. Notice that throughout this paper, we only work
with {\em non-negative} and {\em monotonically non-decreasing}
submodular functions.

\begin{definition}
Let $U$ be the ground set. Let $f(\cdot): 2^{|U|}\rightarrow \real$ be a
function mapping any subset of $U$ to a real number. $f(\cdot)$ is a
submodular function if:
$$\forall S, T\subseteq U, \,\, f(S)+f(T) \geq f(S\cup T) + f(S\cap T).$$
\end{definition}

For simplicity, for any set $S\subseteq U$, we define its marginal function value $f_S(\cdot)$ as
follows. For any $T\subseteq U$, $f_S(T) = f(S\cup T) - f(S)$. For
singletons, we also write $f_S(e) = f_S(\{e\})$.
It is
not difficult to see that $f_S(\cdot)$ is submodular if $f(\cdot)$ is
submodular.

\subsection{The simulated greedy algorithm}


%
Our general algorithm is
based on the greedy algorithm, as in
Algorithm~\ref{alg:general greedy}. 
\begin{algorithm}[h!]
\SetAlgoLined
\caption{$\greedy$}
\label{alg:general greedy}
\KwIn{Set $H \subseteq U$ of matroid $(U, \mi)$ and function $f(\cdot)$}
\KwOut{A set of elements $T\subseteq H$ and $T\in \mi$}
$T \leftarrow\emptyset$\;
\While {$\exists \, e^*= \argmax_{e\in H} \,\{ f_T(e) \,\mid \, T\cup
 \{e\} \in \mi\}$}
{
$T\leftarrow T\cup \{e\}$; $H\leftarrow H\setminus \{e\}$\;
}
return $T$\;
\end{algorithm}

\begin{algorithm}[h]
\caption{$\online$}
\label{alg:general online}
\KwIn{Matroid $(U, \mi)$ and function $f(\cdot)$}
\KwOut{Selected elements $\alg$}
$M,N,\alg \leftarrow\emptyset$\;
$m \leftarrow{}Binom(|U|,p)$\;
Observe the first $m$ elements  $H$\;
$M \leftarrow \greedy(H)$\;
\For{any subsequent element $e$}
{
    \If{$\greedy(H\cup \{e\}) \neq \greedy(H)$} 
    {
          $N\leftarrow N\cup \{e\}$\;
          \If{$\alg \cup \{e\} \in \mi$}
          {
                Accept $e$ and $\alg \leftarrow \alg\cup \{e\}$\;
          }
     } 
}
\end{algorithm}
%
\begin{algorithm}[h]
\caption{$\sim$}
\label{alg:general simulate}
\KwIn{Matroid $(U, \mi)$ and function $f(\cdot)$}
\KwOut{Selected elements $S$}
$H,M,N,S\leftarrow\emptyset$\;
\For{ each element $e$}
{
	Flip a coin with prob.~$p$ of head\; 
	{\bf if} $head$, $H\leftarrow H\cup \{e\}$\;
}
\While {$\exists \, e^*= \argmax_{e\in U\setminus \{M\cup N\}} \,\{ f_M(e) \,\mid \, M\cup
 \{e\} \in \mi\}$}
{
 {\bf if} $e\in H$ {\bf then} $M\leftarrow{}M\cup\{e\}$\; {\bf else} { $N\leftarrow{}N\cup\{e\}$}\;
}
Prune $N$ to produce a set of elements $S\in \mi$\;
\end{algorithm}

Our {\em simulated greedy} algorithm $\online$ works as follows. (We
will discuss the name of {\em simulated greedy} in a minute.)
We observe the first
$m$ elements $H$ without any selection, where $m$ is sampled from
Binomial distribution $Binom(n, p)$ for some chosen probability $p$. Then we compute the greedy
solution $\greedy(H)$. After that, for any subsequent element $e$, we
test that whether the greedy solution will change if $e$ is added to
$H$ hypothetically. If so, we mark $e$ as a candidate and place it in
$N$. Furthermore, if $\alg\cup \{e\} \in \mi$  for candidate $e$ and current $\alg$, we accept $e$ into $\alg$. (Both $N$ and $\alg$ are initially empty.)
The final $\alg$ will be the output of our algorithm. Observe that maintaining set $N$ is not necessary because $N$ only collects elements that has passed the greedy check and might be accepted potentially. However, we keep the notation in the algorithm because it corresponds to the same $N$ in $\sim$, which is heavily used throughout the analysis. 

As we mentioned earlier, $\online$ is a generalization of
the algorithms in~\cite{DimitrovP08,KorulaP09}. In particular, it has
been observed that a {\em simulated} random algorithm in Algorithm~\ref{alg:general
 simulate} can be used in
analyzing the performance of $\online$. (We name $\online$ as a {\em
 simulated greedy} algorithm because of the corresponding greedy
algorithm which simulates the online version.)

More specifically, $\sim$
works as follows. We maintain two sets $M$ and $N$ which are initially
empty. In each step, we select an element $e \in U\setminus (M\cup N)$ such that $f_{M}(e)$ is
maximized and $M\cup \{e\} \in \mi$. (If no such element exists,
$\sim$ terminates.) Then we toss a biased random coin with probability $p$ to be head, which is the same probability in sampling $m$ in $\online$. If the coin is
head, $e$ is placed into $M$. Otherwise, $e$ is placed into $N$.
Since $N$ may not be an independent set in $\mi$ after $\sim$ terminates, we
prune $N$ to produce $S\subset N$ such that $S\in \mi$. The actual pruning
step might be different in different application settings. 

$\sim$ is useful in analyzing the performance of $\online$ with random
arriving elements, because,
as the naming suggests, both $M$ and $N$ have the same joint
distribution in the two algorithms. 
This connection is extensively discussed in~\cite{DimitrovP08,KorulaP09}. 
%
For completeness, we provide a proof
in Appendix~\ref{app:pre}. We will
guarantee that $S$ in $\sim$ is stochastically dominated by $\alg$ in
$\online$. Since we assume $f(\cdot)$ is non-decreasing, in analyzing the performance of $\online$, we can focus
on $S$ in $\sim$. 

\def\lemidentical{
The sets of elements of $H$, $M$ and $N$ by $\sim$ have the same
joint distribution as the $H$, $M$ and $N$ generated by $\online$ with
a random permutation of the elements in $U$.}
\begin{lemma}
\label{lem:identical}
\lemidentical
\end{lemma}
\section{A simple stochastic process}
\label{sec:uniform}
In this section, we study a simple stochastic process which serves as
a building block of our analysis. We will apply this process to either the entire ground set $U$ or some subsets of the elements in $U$. Therefore, although we use the same notation for $M$ and $N$ in this section, they can be viewed as the intersections between the set of elements that are under consideration and the actual global $M$ and $N$ generated by the algorithm.

The simple stochastic process is
defined by an underlying Bernoulli process, with an infinite sequence
of independent and identical random variables $X_{t}\in \{0,1\}$ for
$t\geq 1$. Each variable $X_t$ is a Bernoulli random variable with
probability $p$ to be $1$. 

Our stochastic process is parametrized by a constant $\mu \geq 1$.
We maintain two sets $M$ and $N$, which are initially empty, as
follows. Starting from $t=1$, if $X_t =1$, we place $t$ into $M$;
otherwise, $t$ is placed in $N$. The process immediately terminates
{\em after} $|M| = \mu$. 

We associate a non-negative weight $w_t$ to every time stamp $t$. In particular
$w_t$ is a mapping from the previous $t-1$ random variables $\{X_1,X_2,\ldots, X_{t-1}\}$ to a
non-negative real
number. ($w_1$ is constant by definition. If the process has been terminated before time $t$, we set $w_t =0$.) For any set $T \subseteq \mathbb{N}$,
we define the weight as,
\begin{equation}
w(T) = \sum_{t\in T} w_t(X_1,X_2,\ldots, X_{t-1}).
\end{equation}
Define $w(\emptyset) = 0$. The following proposition shows that the total weights of $M$ and $N$ are close to each other.

\def\proweightofM{$\E[w(M)] = \frac{p}{1-p}\E[w(N)].$}
\begin{proposition}
\label{pro:weight of M}
\proweightofM
\end{proposition}
\begin{proof}
Due to linearity of expectation, it is sufficient to consider the
weights of $M$ and $N$ on a particular time stamp $t$. Let $\cF_t$ be
the $\sigma$-algebra encoding all the randomness up to the time
$t$. Notice that $w_t$ is $\cF_t$-measurable. 
Let $w_t^M = w_t$ if $X_t=1$ and $0$ otherwise. Similarly, we
define $w_t^N =w_t$ if $X_t =0$ and $0$ otherwise. We immediately have
$\E[w_t^M\mid \cF_t] = \frac{p}{1-p} \E[w_t^N\mid \cF_t]$. Therefore,

\[\E[w(M)] = \sum_{t\geq 1} \E_{\cF_t}\left[ \E[w_t^M\mid \cF_t] \,\right] = \frac{p}{1-p}
\sum_{t\geq 1} \E_\cF\left[ \E[w_t^N \mid \cF_t]\, \right] = \frac{p}{1-p}\E[w(N)].\]
\end{proof}

Notice that after the process terminates, we have $|M|=\mu$. On the other hand, the size
of $N$ might be very large. Our analysis will be based on $N$s that
are with size at most $\mu$. We produce an independent set $S$ from $N$ by a pruning process
as follows.


{\bf Pruning.} More formally, to address the issue of too large $N$s,
we define $S = N$ if $|N| \leq \mu$ and $S =
\emptyset$ otherwise. Clearly, we have $S\subseteq N$
and $w(S) \leq w(N)$. 

We want to show that $w(S)$ is close to $w(N)$ in expectation. However , it is not
possible for arbitrary set of 
$\{w_t\}$. In what follows, we impose a ``decreasing
weight'' condition on $\{w_t\}$, which always holds in our applications. This condition
is crucial in building the connection between $w(S)$ and $w(N)$.

\begin{definition}[Decreasing weight mappings]
The set of mappings $\{w_t\}$ forms a sequence of decreasing weight
mappings if for any $i <j$ and $x_1,x_2,\ldots, x_{i-1}, x_i,\ldots,
x_{j-1}$ we have:
\[ w_i(x_1,\ldots, x_{i-1}) \geq w_j(x_1,\ldots, x_{i-1},\ldots, x_{j-1}).\]
\end{definition}

Proposition~\ref{pro:weight of S} makes a connection between $w(S)$ and $w(N)$.
We briefly discuss the intuition behind this statement. Our objective is to show that the weight pruned from $N$ to $S$ is small. The random process indicates that the probability for having a large $N$ is exponentially decreasing on its size, e.g., by the Chernoff bound. Therefore, the probability mass of $N$ that is pruned is small. In terms of weight, on the other hand, those larger $N$s do have greater weights. 

The condition of the decreasing weight mappings comes to rescue. In particular, in this case,
the weight of $N$ grows roughly ``linear'' to its size. As the probability decreases exponentially with the size of $N$, the total weight pruned can still be bounded as the summation of a geometric sequence for those large $N$s. We concretely implement this proof as follows.

\def\proweightofS{
Let $\beta=2e(1-p)$. If $\{w_t\}$ forms a sequence of decreasing
weight mappings, we have
 \[ \E[w(N)]-\E[w(S)]\leq
 \frac{(\mu+1-\mu\beta)\beta^\mu}{(1-\beta)^2}\cdot\E[w(S)]
\leq \frac{(\mu+1-\mu\beta)\beta^\mu}{(1-\beta)^2}\cdot\E[w(N)]
\]
If $\mu =1$, it can be improved to 
\[ \E[w(N)] - \E[w(S)] \leq \frac{1-p^2}{p^2} \cdot \E[w(S)] \leq
\frac{1-p^2}{p^2} \cdot \E[w(N)]. \]}
\begin{proposition}
\label{pro:weight of S}
\proweightofS
\end{proposition}
\begin{proof}
To simplify the notation, let $h_N = |N|$.
By definition:
  \[\E[w(N)]=\sum_{k=1}^{\infty}\E[w(N)\mid h_N=k]\pr[h_N=k] ;\, 
\E[w(S)]=\sum_{k=1}^{\mu}\E[w(N)|h_N=k]\pr[h_N=k]\]
  Let $N_i$ be the set of the first $i$ elements of $N$
in our stochastic process.
Let $\cA_k$ be all possible outcomes of $N_k$. 
  Now for any fixed $1\le k\leq \mu < q$, by definition, we have
\begin{equation}
\label{eqn:wnk}
\E[w(N_k)\mid h_N=q] \pr[ h_N=q]=\sum_{A\in\cA_k}w(A)\cdot\pr[N_k=A\wedge{}h_N=q].
\end{equation}

For a fixed $A$, let $\ell(A)$ be the number of $1$s in $X_{t}$s
 when we pick the last element in $A$. (The last one must be $0$ as
 it goes into $N$.) 
Since $A\in \cA_k$, $\ell(A) < \mu$. We have
  \begin{align*}
    \frac{\pr[N_k=A\wedge{}h_N=q]}{\pr[N_k=A\wedge{}h_N=k]}=&\frac{\pr[h_N=q \mid N_k=A]\pr[N_k=A]}{\pr[h_N=k \mid N_k=A]\pr[N_k=A]}\\
    =&\frac{\binom{k+\ell(A)-1}{\ell(A)}p^{\ell(A)}(1-p)^k \cdot \binom{q-k+\mu-\ell(A)-1}{q-k}p^{\mu-\ell(A)}(1-p)^{q-k}
    }{
    \binom{k+\ell(A)-1}{\ell(A)}p^{\ell(A)}(1-p)^k \cdot p^{\mu-\ell(A)}
    }\\
    =&\binom{q+\mu-\ell(A)-k-1}{q-k}(1-p)^{q-k}\\
   \leq& \binom{q+\mu-2}{q-1} (1-p)^{q-k}\leq (2e)^{q-1}(1-p)^{q-k}
  \end{align*} 
  The first inequality comes from the fact that
  $\binom{n-t}{k-t}\leq \binom{n}{k}$ and $\binom{n-t}{k} \leq
  \binom{n}{k}$ when $t\geq 0$. The last inequality is due to $\binom{n}{k}\le
  (\frac{ne}{k})^k$ and $\mu < q$. So we have, with Eqn.(\ref{eqn:wnk}),
  \begin{equation}
  \label{eqn:qtok}
  \E[w(N_k)\mid  h_N=q] \pr[ h_N=q]\le(2e)^{q-1}(1-p)^{q-k}\E[w(N_k)
    \mid h_N=k] \pr[h_N=k]
  \end{equation}
  On the other hand, by the decreasing order of $w_i$, we have that
  \begin{equation}
\label{eqn:qk}
\E[w(N)\mid h_N=q]\le\frac{q}{k}\E[w(N_k)\mid h_N=q].
\end{equation}
  Therefore, for any $q > \mu$, 
  \begin{align*}
    \E[w(S)]&=\sum_{k=1}^\mu{\E[w(N_k)\mid h_N=k] \pr[h_N=k]}\\
    &\ge\sum_{k=1}^\mu(2e)^{1-q}(1-p)^{k-q}\E[w(N_k)\mid h_N=q] \pr[h_N=q]   \mbox{ \ \ \ \ \ \ \ \   by Eqn.(\ref{eqn:qtok})    }\\
    &\ge\sum_{k=1}^\mu\frac{k}{q}(2e)^{1-q}(1-p)^{k-q}\E[w(N)\mid h_N=q]
    \pr[h_N=q] \mbox{ \ \ \ \ \ \ \   by Eqn.(\ref{eqn:qk})    }\\
    &=\frac{(1-p)^{-q}}{(2e)^{q-1}q}\cdot\E[w(N)\mid h_N=q]\pr[h_N=q]\cdot\sum_{k=1}^\mu
    k(1-p)^k\\
    &\ge\frac{(1-p)^{-q}}{(2e)^{q-1}q}\cdot\E[w(N)\mid h_N=q]\pr[h_N=q]\cdot(1-p)
  \end{align*}
  Finally, recall that $\beta=2e(1-p)$, we have
  \begin{align*}
\E[w(N)]-\E[w(S)]
& = \sum_{q=\mu+1}^{\infty}\E[w(N)\mid h_N=q]\pr[h_N=q]\\
    &\le \E[w(S)]\sum_{q=\mu+1}^{\infty}q(1-p)^{q-1}(2e)^{q-1}\\
    &= \E[w(S)]\cdot\frac{(\mu+1-\mu\beta)\beta^\mu}{(1-\beta)^2}
   \end{align*}
  The last equality come from the fact that, for any $\alpha<1$,
  $\sum_{i=k}^{\infty}i\cdot\alpha^i=\frac{\alpha^k(k+\alpha-k\alpha)}{(1-\alpha)^2}$.

Now consider the case that $\mu = 1$. By the stochastic process, $w(N_1)$
is either $0$ or $w_1$.
Eqn.~(\ref{eqn:qk}) still holds. 
  \begin{align*}
    \E[w(S)]&={\E[w(N_1)\mid h_N=1] \pr[h_N=1]}\\
    &= \E[w(N_1)\mid h_N=q] \pr[h_N=q] (1-p)^{1-q}\\
    &\ge \frac{1}{q}(1-p)^{1-q}\E[w(N)\mid h_N=q]
    \pr[h_N=q]
  \end{align*}

  \begin{align*}
\E[w(N)]-\E[w(S)]
& = \sum_{q= 2}^{\infty}\E[w(N)\mid h_N=q]\pr[h_N=q]\\
    &\le \E[w(S)]\sum_{q=2}^{\infty}q(1-p)^{q-1}\\
    &= \frac{1-p^2}{p^2}\E[w(S)].
   \end{align*}
\end{proof}

\section{Laminar Matroid}
\label{sec:laminar}
In this section, we study the performance of our simulated greedy algorithm $\sim$ for the submodular matroid secretary problem
with a laminar matroid. 
We first show that the entire process of $\sim$ can be casted as a simple stochastic process as discussed in the previous section. After that, we inspect the pruning stage in details. In particular, for each $B_i$ in the laminar matroid, we study a simple stochastic process restricted on the elements in $B_i$. The loss on the entire pruning steps can be divided into losses on the $B_i$s, which can be bounded by Proposition~\ref{pro:weight of S}.

Let $\mu$ be the rank of the laminar matroid. 
Essentially, $\sim$ will
select (at most) $\mu$ elements. We cast
the $\sim$ process to the simple stochastic process with $\mu$ as
follows. 

In the $t$-th round, when the first $t-1$ random coins are
tossed, the current element $e$ in the greedy order is uniquely
defined, as well as the current $M$ and $N$. We define the weight
$w_t = f_{M_e}(e)$ where $M_e$ is the current elements in $M$.

{\bf Remark.} We make two remarks regarding the connection
between the two stochastic processes. First, the original simple stochastic process terminates
when $|M| =\mu$. $\sim$
might terminate earlier because of the limit on the number of
elements. In such cases, we assume the availability of an {\em infinite}
number of dummy elements, with zero weights, which will eventually fill up $M$. In particular, when any of
these dummy element arrives at time $t$, $w_t=0$ with respect to the previous
random outcomes. Notice that these dummy elements will enlarge the size of $N$ without
increasing the weights of $N$ and $S$. So all conclusions we draw in
last section still hold. Second, $M$ (as well as $N$ and $S$) in the simple stochastic process
consists of time stamps, while in all processes we study later $M$
consists of real elements. Nevertheless, for every real element $e\in
M$, we define $w(e) = w_t$ where $t$ is the time $e$ appears in the
greedy order of $\sim$. Both
$w_t$ and $w(e)$ are random variables. We have
$w(M) = \sum_{e\in M} w(e)$.

We extend the $w(\cdot)$ to elements besides those in $M$.
In particular, $w(e) = f_{M_e}(e)$ for $e\in M\cup
N$, i.e., $e$ appears in the greedy order of $\sim$, where $M_e$ is the
current set of elements in $M$ when $e$ appears. If $e\notin M\cup N$, set $w(e) = 0$. Notice that $w(M) =
f(M)$ by definition. 
%
Furthermore, each element in the offline optimal solution has
probability $p$ in $H$, i.e., a head coin is associated with it. 
By
submodularity of $f(\cdot)$, the expected value of the optimal solution in $H$ is
at least $p\cdot \opt$. On the other hand, the greedy algorithm is a
$2$-approximation with a matroid constraint when the valuation function is monotone and submodular. Together with Proposition
\ref{pro:weight of M}, we have 
\begin{lemma}
\[\E[f(M)]=\E[w(M)]=\frac{p}{1-p}\E[w(N)] \geq \frac{p}{2} \cdot \opt\]
\end{lemma}

{\bf Pruning.} Notice that although $M$ is independent, $N$ might
not be independent. We obtain $S$ by pruning $N$ as follows.

\begin{equation}
\label{eqn:laminar S pruning}
S=N \setminus \left(\bigcup_{B\in \mathcal{F}} \one_{|N\cap
   B|>\mu(B)}\cdot (N\cap B)\right),
\end{equation}
where $\one_{cond}\cdot (N\cap B) = N\cap B$ if $cond$ is true and
empty otherwise. In other words, if one constraint $B_i$ is violated in
$N$, we remove all elements in $B_i$ from $N$. Clearly, $S$ is
independent. Furthermore, since $\alg$ will be the greedy independent
set of $N$ for a random order, it is straightforward to show that
$S\subseteq \alg$.

%
Therefore, it is sufficient to bound $\E[f(S)]$. To do that, we first
provide a lower bound for 
$\E[w(S)]$. After that, we bound $\E[f(S)]$ in terms of $\E[w(S)]$.

{\bf Roadmap.}
Here we briefly outline our strategy in getting the two pieces of
results. To measure $\E[w(S)]$, we estimate the weight loss due to
the pruning in Eqn.(\ref{eqn:laminar S pruning}). For each
constraint $B_i$, we cast the stochastic process in $\sim$ in processing elements in $B_i$
into a simple stochastic process with $\mu(B_i)$. By invoking
Proposition~\ref{pro:weight of S}, the weight loss $w(N\cap B_i) -
w(S\cap B_i)$ is $2^{O(\mu(B_i))}\cdot w(N\cap B_i)$,
which is charged to all elements in $B_i$ proportionally to
$\one_{e\in N} w(e)$ for all $e\in B_i$. The catch here is, for each element
$e\in U$, the set of $\{B_i\}$ containing $e$ has a strictly
increasing $\{\mu(B_i)\}$ sequence. Therefore, the charges on $e$ form a geometric
sequence which in total will not exceed a constant fraction of
$\one_{e\in N}\cdot w(e)$. Since $w(N) = \sum_{e\in N} w(e)$, the
total weight loss is a constant fraction.

The second piece of ingredient is to make a connection between
$\E[f(S)]$ and $\E[w(S)]$. For simplicity, let us consider $\E[f(N)]$
and $\E[f(M)]$ instead to convey the idea. Recall that
$w(N) = \sum_{e\in N} f_{M_e}(e)$, where $M_e$ is the set of elements
in $M$ when $e$ arrives. Therefore, it is not intuitive why
$\E[f(N)]$ should be large in the first place. To elaborate, we
consider function $F = f(M) +2f(N) - f(M\cup N)$ during the execution of
the algorithm, which is a lower bound of $2f(N)$. We can view $f(M)+f(N)-f(M\cup N)$ as the {\em intersection} between $M$ and $N$, e.g., if
$f(\cdot)$ is modeling a set cover. During the execution of the
algorithm, when $e$ arrives, we have two cases: (1) $f_{M_e}(e) \approx f_{N_e}(e)$, where
$M_e$ and $N_e$ are the current set of $M$ and $N$ respectively. $F$
will grow nicely proportional to $f_{M_e}(e)$ in this case. (2) $f_{M_e}(e) \gg
f_{N_e}(e)$. Notice $e$ is placed into $M$ with
probability $p$, in which case $F$ grows proportional to $f_{M_e}(e)$
as well. This is because $f_{M_e\cup N_e}(e) \leq f_{N_e}(e) \ll
f_{M_e}(e)$ due to the submodularity of $f(\cdot)$. Therefore, $F$
grows in both cases in expectation, which gives a lower bound for
$\E[f(N)]$ with respect to $\E[f(M)]$. The analysis in bounded $f(S)$ is more complicated. Though
the underlying idea is identical. We formally implement these two ideas in Lemma~\ref{lem:laminar w(S)} and 
Lemma~\ref{lem:laminarS}.

\begin{lemma} 
\label{lem:laminar w(S)}
Let $\beta = 2e(1-p)$. We have
\[\E[w(S)]\ge(1-\frac{2\beta}{(1-\beta)^3})\E[w(N)].\]
\end{lemma}
\begin{proof}
%
%
Since for a fixed set of random outcomes, $w(\cdot)$ is a linear
function. 
By Eqn.(\ref{eqn:laminar S pruning}), we have that
%
\[\E[w(N)] \le \E[w(S)] + \sum_{B\in \mathcal{F}}\E[w(\one_{|N\cap
    B|>\mu(B)}\cdot (N\cap B))]. \]

Now we focus on the term $\E[w(\one_{|N\cap{}B|>\mu(B)}\cdot (N\cap B))]$ and the
simulated greedy algorithm on elements in $B$, i.e., a particular
constraint in $\cF$. We isolate $B$ in the
process by rearranging the randomness
as follows. First, for each element in $U\setminus B$, we assign an independent
random coin to it, i.e., if this element appears in the algorithm, its
random coin will be tossed. For a fixed outcome of all random coins
outside of $B$, the simulated greedy algorithm is a {\em simple stochastic process} for
the elements in $B$. The only difference, however, is the process may
terminate before $|M\cap B| = \mu(B)$. This can be easily remedied by
appending dummy elements as before.
Recall that
$\beta=2e(1-p)$. 
By Proposition \ref{pro:weight of S}, we have:


\begin{equation}
\E[\one_{|N\cap{}B|>\mu(B)}\cdot w(N\cap B)] \le \frac{(\mu(B)+1-\mu(B)\beta)\beta^{\mu(B)}}{(1-\beta)^2}\cdot \E[w(N\cap
B)].
\end{equation}

 It follows that
\begin{eqnarray}
\E[w(N)] &\le &\E[w(S)] + \sum_{B\in\cF}\frac{(\mu(B)+1-\mu(B)\beta)\beta^{\mu(B)}}{(1-\beta)^2}\cdot \E[w(N\cap B)] \nonumber\\
&=& \E[w(S)] + \frac{1}{(1-\beta)^2}\sum_{B\in\cF}\sum_{e\in U}\E[(\mu(B)+1-\mu(B)\beta)\beta^{\mu(B)}\cdot w(e)\one_{e \in B} \cdot
\one_{e\in N}]\nonumber\\
&=& \E[w(S)] + \frac{1}{(1-\beta)^2}\sum_{e\in U}\E\left[w(e)\one_{e\in N}\left(\sum_{B\in\cF}(\mu(B)+1-\mu(B)\beta)\beta^{\mu(B)}\cdot\one_{e \in B}\right)\right] \nonumber\\
&\le& \E[w(S)] + \frac{1}{(1-\beta)^2}\sum_{e\in U}\E[ w(e)\one_{e\in N}]\left(\sum_{i\ge 1}(i+1-i\beta)\beta^i\right) \label{eqn:geometric sequence}\\
&=& \E[w(S)] + \frac{2\beta}{(1-\beta)^3}\E[w(N)] \nonumber
\end{eqnarray}
Eqn.(\ref{eqn:geometric sequence}) uses the fact that the set of constrains $\{B_i\}$ containing an element $e$ has a strictly increasing sequence of $\{\mu(B_i)\}$.
%
\end{proof}

We then bound $\E[f(S)]$ as follows. For an element $e$, let $N_e$ be
the set of elements in $N$ when $e$ appears in $\sim$. We define
$g(e) = f_{N_e}(e)$ if $e \in M\cup N$ and $g(e) = 0$
otherwise. 
\footnote{We define $g(e)$ based on $N_e$ instead of $S_e$,
  i.e., the current set of elements in $S$, because $S_e$ is still a
  random set even all the randomness before $e$'s arrival is fixed.}

\begin{lemma}
\label{lem:laminarS}
For any $t>0$, 
let $\theta = 1 +\frac{(1-p)t}{p}$.
We have
\[\E[f(S)]\geq(\frac{1}{\theta}-\frac{(1-\beta)^3}{t((1-\beta)^3-2\beta)})\E[w(S)]\]
\end{lemma}
\begin{proof}
Let $g(S)=\sum_{e\in S}g(e)$. Since
$S\subseteq N$, we have $f(S)\geq g(S)$ by the submodularity of $f(\cdot)$. We inspect the function
$F(S,M,N)=t\cdot g(S)+f(M)-f(M\cup N)$. By the monotonicity of $f$,
$f(S) \geq g(S)\geq F(S,M,N)/t$.

Define $\Delta_e=F(S_e',M_e',N_e')-F(S_e,M_e,N_e)$ where $M'_e$ (resp.
$N'_e$ and $S'_e$) is the
set $M$ (resp. $N$ and $S$) after we process element $e$.
If $e\notin M\cup N$, define $\Delta_e = 0$.
Therefore, $F(S,M,N)=\sum_{e\in U}\Delta_e$. Let $\cR_e$
be the sub-$\sigma$-algebra encoding all randomness up to the time
$e$ is picked in $\sim$. Notice that $M_e$ and $N_e$ are $\cR_e$
measurable. We have $\pr[e\in M\mid \cR_e]=p$ and $\pr[e\in
N\mid \cR_e]=1-p$.
\begin{align*}
  \E[\Delta_e\mid \cR_e]&=t\cdot(\E[g(S')-g(S)\mid
  \cR_e])+(\E[f(M')-f(M)\mid \cR_e])\\
  &-(\E[f(M'\cup N')-f(M\cup N)\mid \cR_e])\\
  &=t\cdot\pr[e\in S\mid \cR_e]f_{N_e}(e)+\pr[e\in M\mid \cR_e]f_{M_e}(e)-f_{M_e\cup N_e}(e)
\end{align*}
Then we bound $\E[\Delta_e\mid \cR_e]$ by case analysis. Notice that $\pr
[e\in M \mid \cR_e] + \Pr[e\in N\mid \cR_e] = 1$ and $\pr[e\in N\mid
\cR_e] \geq \pr[e\in S \mid \cR_e].$ \\
Case 1: $f_{M_e}(e)\geq \theta\cdot f_{N_e}(e)$.
\begin{align*}
\E[\Delta_e\mid \cR_e]&\geq\pr[e\in M\mid \cR_e](f_{M_e}(e)-f_{M_e\cup
  N_e}(e))-\pr[e\in N\mid \cR_e]f_{M_e\cup N_e}(e)\\
&\geq \frac{p}{1-p} (1 - \frac 1\theta)\Pr[e\in  S\mid \cR_e] f_{M_e}(e)-\pr[e\in N\mid\cR_e]f_{{M_e}}(e)
\end{align*}
Case 2: $f_{M_e}(e)< \theta\cdot f_{N_e}(e)$.
\begin{align*}
\E[\Delta_e\mid\cR_e]&\geq t\cdot\pr[e\in S\mid
\cR_e]f_{N_e}(e)-\pr[e\in N\mid \cR_e]f_{M_e\cup N_e}(e)\\
&\geq \frac{t}{\theta}\pr[ e\in S\mid \cR_e] f_{M_e}(e)-\pr[e\in N\mid
\cR_e]f_{{M_e}}(e)
\end{align*}
By definition of $\theta$, we have $\frac{p}{1-p}(1 - \frac 1\theta)=t/\theta$. So
\[\E[\Delta_e\mid \cR_e]\geq\frac{t}{\theta}\pr[e\in
S\mid \cR_e]f_{M_e}(e)-\pr[e\in N\mid \cR_e]f_{{M_e}}(e)\]
Therefore
\begin{align}
  t\cdot \E[f(S)]&\geq
  \E[F(S,M,N)]=\sum_e{\E_{\cR_e}[\E[\Delta_e\mid \cR_e]]} \nonumber\\
  &\geq\sum_{e\in U}{\E_{\cR_e}[\frac{t}{\theta}\pr[e\in
    S\mid \cR_e]f_{M_e}(e)-\pr[e\in N\mid \cR_e]f_{{M_e}}(e)]}\nonumber\\
  & = \frac{t}{\theta}\E[ w(S)]-\E[ w(N)] \label{eqn:savedforlater}\\
  &\geq\frac{t}{\theta}\E[w(S)]-\frac{(1-\beta)^3}{(1-\beta)^3-2\beta}\E[
  w(S)] \nonumber
\end{align} 
The last inequality is by Lemma~\ref{lem:laminar w(S)}. 
So $\E[f(S)]\geq(\frac{1}{\theta}-\frac{(1-\beta)^3}{t((1-\beta)^3-2\beta)})\E[w(S)]$
\end{proof}

Combining all the results together, we have an algorithm with
competitive ratio at most $211$ with $p = 0.9794$ and $t=10.1415$. 
\begin{theorem}
There is an online algorithm with competitive ratio at most $211$ for
the submodular matroid secretary problem with laminar matroids.
\end{theorem}



\subsection{The intersection of constant number of laminar matroids}

\begin{theorem}
  \label{thm:intersection}
  For any constant $k$, there is an online algorithm with competitive
  ratio at most $\frac{1000k(k+1)}{9}$ for the submodular matroid
  secretary problem with the intersection of $k$ laminar matroids.
\end{theorem}
\begin{proof}
The independent set  we considered is the intersection of $k$
matroids. Therefore, the greedy algorithm is a $1/(k+1)$-approximation. 
Together with Proposition~\ref{pro:weight of M}, 
we have
\begin{equation}
\label{eqn:intersectionwn}
\E[f(M)] =\E[w(M)] = \frac{p}{1-p}\E[w(N)] \ge \frac{p}{k+1}\cdot \opt.
\end{equation}

Following the proof of Lemma~\ref{lem:laminar w(S)}, we have
\begin{equation}
\label{eqn:intersectionws}
\E[w(S)]\ge \left(1-k\cdot\frac{2\beta}{(1-\beta)^3}\right)\E[w(N)],
\end{equation}
where $\beta = 2e(1-p)$.
The additional $k$ terms come from the fact that we have to sum up $k$ geometric sequences instead of 
one in Eqn.(\ref{eqn:geometric sequence}).

Let $\gamma = 1-k\cdot\frac{2\beta}{(1-\beta)^3}$. 
We can follow the proof of Lemma~\ref{lem:laminarS} until the inequality of Eqn.(\ref{eqn:savedforlater}), which remains true. In particular, for any $t>0$,
\[t\cdot \E[f(S)]\ge \frac{t}{\theta}\E[ w(S)]-\E[ w(N)]\, \Rightarrow \,
\E[f(S)]\ge (\frac{1}{\theta}-\frac{1}{t\gamma})\E[ w(S)],\]
Let $a = \frac{1-p}{p}$. Recall that $\theta = 1 + \frac{1-p}{p}\cdot t  =  1 + at$. 
By taking $t = \frac{1}{\sqrt{a}(\sqrt{\gamma}-\sqrt{a})}$. 
We have that 

\begin{equation}
\label{eqn:intersectionfsws}
\E[f(S)]\ge \frac{(\sqrt{\gamma}-\sqrt{a})^2}{\gamma}\E[ w(S)]
\end{equation}

Thus overall we have that 
\begin{align*}
\E[f(S)] &\ge (\sqrt{\gamma}-\sqrt{a})^2 \E[ w(N)]\,\,\,\,\,\,\,\,\,\, &\mbox{   /* by    Eqn.(\ref{eqn:intersectionws}) and Eqn.(\ref{eqn:intersectionfsws}) */}\\
 &\ge (\sqrt{\gamma}-\sqrt{a})^2 \frac{1-p}{k+1} \cdot \opt &\mbox{/*  by Eqn.(\ref{eqn:intersectionwn})*/ }
\end{align*}

Now we analyze this ratio. Set $p =  1 - \frac{c}{k}$ for some sufficiently small constant $c$. Then $\beta = 2ec/k \le 2ec < 1$ and $\gamma =  1 - \frac{4ec}{(1-\beta)^3}\ge 1 - \frac{4ec}{(1-2ec)^3}$. 
By enforcing $c<0.04$, we have $\gamma > a$.
Then 
\[(\sqrt{\gamma}-\sqrt{a})^2 \frac{1-p}{k+1} \ge \frac{1}{k(k+1)}\cdot c\left(\sqrt{1 - \frac{4ec}{(1-2ec)^3}} - \sqrt{\frac{c}{1-c}}\right)^2\]

By taking $c = 0.02$. We have that 
\[(\sqrt{\gamma}-\sqrt{a})^2 \frac{1-p}{k+1} \ge 0.009\frac{1}{k(k+1)}\]

\end{proof}

\subsection{The linear case}
In this section, we analyze the algorithm $\online$ for
the laminar matroid secretary problem with linear functions. For
this special case, we improve the competitive ratio to $9.6$.

\begin{theorem}\label{thm:app:laminar_linear}
  Algorithm \ref{alg:general online} is a 9.6-competitive algorithm
  for the linear matroid secretary problem with laminar matroids.
\end{theorem}

For linear functions, our main idea is to prove that each
element in the optimal solution has a good probability of staying in
our solution set $S$ in $\sim$. Before proving the theorem directly, we first
define some useful random variables and analyze the random process used
in $\sim$ more precisely.

\begin{definition}
  Let $X_1,X_2,\ldots,X_n$ be independent Bernoulli trials such that
  $\pr[X_i=1]=p$ and $\pr[X_i=0]=1-p$. Define $i_0^X(k)$ to be the
  random variable indicating the index of the $k^{th}$ appearance of
  $0$'s in the sequence, $i^X_1(k)$ that of $k^{th}$ appearance of
  $1$'s. We define $i_0^X(0)=i_1^X(0)=0$. Define $G_p(m,n)$
  for any positive integer $m,n$ as follows.
  \begin{equation*}
    G_p(m,n) = \Pr[i^X_1(m) > i^X_0(n)]
  \end{equation*}
\end{definition}

Intuitively, in $\sim$, we flip a coin for each
element and add it $M$ if and only if the coin is head.
We couple $\sim$ with $\{X_i\}$ as follows. 
If $X_i =1$, the $i$-th element in the greedy order of $\sim$ will be placed into $M$. Otherwise, it is placed into $N$.
Consider the order of elements greedily selected in $\sim$. Then $i_1^X(k)$ (resp. $i_0^X(k)$) can be
viewed as the index of the $k$th element added to $M$ (resp. $N$) in this greedy order. Since all elements
considered in $\sim$ are ordered by weights, $G_p(m,n)$ denotes the probability that 
the weight of the $m$th element in $M$ is smaller than the weith of the $n$th element in $N$.

Consider any element $e$ in the offline optimal solution. In $\sim$, $e$ will be in $M$ if the random coin comes with head when it is processed. Otherwise $e$ will be placed in to $N$. (Since the valuation function is linear, $e$ will always show up in the greedy order in $\sim$.) Therefore, the probability that $e\in N$ is $1-p$. The difficult part is to argue that $e$ will survives the pruning with good probability.

We will use the same pruning process as in Eqn~(\ref{eqn:laminar S pruning}). In the following, we show that for any $B$ that contains $e$, the probability that $B$ is violated, i.e., $\mu(B) < N\cap B$, is at most $G_p(\mu(B),\mu(B))$. In particular, we have the following lemma.

\begin{lemma}\label{lem:app:Gmn}
   \[\forall e\in U,\,\Pr[e\in S \mid  e\in N] \ge 1 - \sum_{n\ge 1}G_p(n,n)\]
\end{lemma}

\begin{proof}
  For any fixed $B$ with $e \in B \in \mathcal{F}$, consider the sequence of
  coins that are tossed in the $\sim$ when the
  elements in $B\setminus \{e\}$ arrive as $\{X_1,X_2,\ldots\}$. (Note that conditioned on
  $e \in N$, we know that the coin toss for $e$ is 0.)

  Conditioned on the event that $e\in N$, the event $|N \cap B| >
  \mu(B)$ implies  $i^X_1(\mu(B)) > i^X_0(\mu(B))$. 
Otherwise, $M \cap B$ will have cardinality $\mu(B)$ before
  $N\cap B$ has cardinality more than $\mu(B)$, and will prevent any
  element in $B$ being added to either $M$ or $N$. (Including $e$, it means that $N$ must reach size $\mu(B)+1$ before $M$ reaches size $\mu(B)$.) That is, $\Pr[|N\cap
  B| >\mu(B) \mid e\in N] \le \Pr[i^X_1(\mu(B)) > i^X_0(\mu(B))] =
  G_p(\mu(B),\mu(B))$. If for each $B \in \mathcal{F}$ that contains
  $e$, we have $|N \cap B| \le \mu(B)$, then $e$ must be in $S$. Thus
  by union bound,
  \begin{eqnarray*}
    \Pr[e\in S \mid e\in N]& \ge& 1 - \sum_{
    B\in
      \mathcal{F} \mid e\in B}\Pr[|N \cap B| > \mu(B) \mid e \in N]
    \\&\ge& 1 - \sum_{B\in \mathcal{F} \mid e\in B}G_p(\mu(B),\mu(B)) \ge 1
    - \sum_{n\ge 1}G_p(n,n)
  \end{eqnarray*}
\end{proof} 

To lower bound the term $\Pr[e\in S \mid e\in N]$, it suffices to upper bound
$G_p(n,n)$ as shown in the following lemma.

\begin{lemma}
\label{lem:gmn}
  \[G_p(m,n) = (1-p)^n \sum_{i = 0}^{m-1}{n-1+i \choose i}p^i \le (1-p)^n(1+p)^{n+m-2}\]
\end{lemma}
\begin{proof}
We prove the first equality by a counting argument. Notice that $G_p(m,n)$ is the probability that when the number of 0s reaches $n$, the number of 1s is still smaller than $m$. 

Let the number of 1s be $i$ before the number of 0s reaches $n$. Then we are interested in $0\leq i\leq m-1$. Consider the first $n+i$ random variables in $X$. Clearly $X_{n+i} = 0$ because this is the time the number of 0s reaches $n$. Therefore, the number of such configurations is $n-1+i \choose i$, each appears with probability $(1-p)^np^i$. The equality comes by summing over all such $i$s.

%
%
%
Finally,
\[G_p(m,n) = (1-p)^n \sum_{i = 0}^{m-1}{n-1+i \choose i}p^{i}\le  (1-p)^n \sum_{i = 0}^{m-1}{m+n-2\choose i}p^{i} = (1-p)^n(1+p)^{m+n-2}.\]
\end{proof}

Finally, we can prove the Theorem \ref{thm:app:laminar_linear} by
showing the following lemma.
\begin{lemma}
  For any element $e\in \opt$, $\Pr[e \in S] \ge (1-p)\left(1 -
    \frac{(1-p)}{1-(1-p)(1+p)^2}\right) \ge 1/9.6$ by taking $p = 0.842$.
\end{lemma}
\begin{proof}
  Since $e\in \opt$, it is straightforward that $e \in M\cup N$, and
  $\Pr[e\in N] = 1-p$. 

\begin{align*}
\Pr[e\in S] & = \Pr[e\in S\mid e\in N]\cdot \Pr[e\in N]\\
& = (1-p) \left(1- \sum_{n\geq 1} G_p(n,n)\right)\\
& \geq (1-p)\left( 1-\sum_{n\ge 1}(1-p)^n(1+p)^{2n-2}\right)\,\,\,\,\, \mbox{by Lemma~\ref{lem:gmn}}\\
& = (1-p)\left(1- \frac{1-p}{1-(1-p)(1+p)^2}\right)
\end{align*}
%
\end{proof}

\section{Transversal matroid}
\label{sec:transversal}


In this section, we apply our simulated greedy algorithm to the submodular matroid secretary problem with transversal matroids. More
specifically, we study the following submodular bipartite
vertex-a-time matching problem.

\subsection{Submodular Bipartite Vertex-a-time Matching Problem}

Korula and P{\'a}l~\cite{KorulaP09} generalized the transversal
matroid secretary problem to an online bipartite graph matching
problem, motivated by~\cite{DimitrovP08}. We further generalize to
submodular valuation functions. In particular, we introduce the {\em
Submodular Bipartite Vertex-at-a-time Matching} (SBVM) problem.

In the SBVM problem, there is an underlying bipartite graph $G(L\cup
R, E)$. We are given the set of right nodes $R$. The nodes in $L$ are arriving sequentially in {\em random order}. When a
vertex $\ell \in L$ arrives, all edges incident to $\ell$ are
revealed. We assume the availability of an oracle for the submodular
valuation function, which we can query the value of any
subset of the edges we have seen. We must immediately decide to accept an
edge to match $\ell$ with an {\em unmatched} vertex of $R$ or drop all
edges incident to $\ell$.

We claim that the matroid secretary problem under a transversal
matroid is a special case of the SBVM problem, when the valuation
function is submodular. In particular, the valuation on $L$ in the
transversal matroid can be extended to the valuation on the edges
$E$. Let $f'(\cdot)$ be the submodular function defined on the subsets of $L$. We
define a function $f(\cdot)$ on the subsets of $E$ as follows: for
$E'\subseteq E$, $f(E') = f'( L \cap E')$, where $L\cap E'$ is the set
of left nodes incident to $E'$.\footnote{The ties in the
valuation function have to be broken in a consistent way.}

\begin{lemma}
\label{lem:exchange_of_submodular}
If $f'(\cdot)$ is a monotonically non-decreasing submodular function, $f(\cdot)$ is
a monotonically non-decreasing submodular function.
\end{lemma}
\begin{proof}
Clearly, if $f'(\cdot)$ is monotonically non-decreasing, $f(\cdot)$ must be
monotonically non-decreasing as well. Let $E''\subseteq E'\subseteq E$. We have
$E''\cap L\subseteq E'\cap L$. Therefore, for any edge $e\in E$, we
want to show that

\begin{equation}
\label{eqn:g}
f(E''\cup \{e\}) -f(E'') \geq f(E'\cap \{e\}) - f(E').
\end{equation}

If $e$ is sharing the left node with $E'$, by monotonicity, the left
term in Eqn.(\ref{eqn:g}) is non-negative while the right term is
zero. So the statement is true. On the other hand, when $e$ is not
sharing the left node with $E'$, $e$ is not sharing the left node with
$E''$ either. Eqn.(\ref{eqn:g}) in this case comes directly from the
submodularity of $f(\cdot)$.
\end{proof}

Hence, for the submodular matroid secretary problem with a
transversal matroid, we can extend the valuation function on $L$ to
the set of edges in the underlying bipartite graph. The optimal
solutions for both problems are the same. In fact, if we find a
matching, which is a good approximation of the SBVM problem, the left
nodes of the matching are a good approximation of the matroid secretary
problem with the same approximation ratio.

Now we are ready to show that our general online algorithm
(Algorithm~\ref{alg:general online}) with slightly modification
gives a constant competitive ratio for the SBVM problem. We
first prove that the greedy algorithm has a good approximation for the offline version of this submodular maximization problem.
\def\lembvm{
For a bipartite graph $G(L,R,E)$ and a monotonically increasing
submodular function $f(\cdot)\geq 0$ defined on all subsets of $E$,
$\greedy$ is a $3$-approximate algorithm.}
\begin{lemma}
\label{lem:bvm}
\lembvm
\end{lemma}
\begin{proof}
First we show that all matchings of $G = (L,R,E)$ can
be represented by independent sets, which are the intersection of two
partition matroids. Both ground sets of these two partition matroids are
$E$. In the matroid $M_1(E, \cI_1)$ (resp. $M_2(E,\cI_2)$), a set of edges is independent if no
two edges in it have the same left (resp. right) node. It is easy to see that
the set of matchings in $G$ is exactly $\cI_1\cap \cI_2$.

The Theorem $2.1$ in \cite{FisherNW78} shows that
the greedy algorithm is a $k+1$ approximation for the submodular
function maximization problem under the intersection of $k$
matroids. Therefore, our algorithm has approximation ratio 3.
\end{proof}

Again, the general online algorithm can be simulated by the following
offline algorithm (Algorithm~\ref{alg:simulate}). We can rearrange the
randomness in $\sim2$ by associating each node in $L$ with a biased
random coin.  In this case, for each edge in $\opt$, with probability
$p$, the coin associated to the left node incident to it is head.
Since $f(\cdot)$ is a submodular function, and $M$ is the greedy
solution for all edges incident to left nodes with associated coins to
be head, we have the following result.
\begin{lemma}
\label{lem:transversal f(M)}
\[ \E[f(M)] \geq \frac{p}{3}\cdot \opt \]
\end{lemma}

\begin{algorithm}
\SetAlgoLined
\caption{$\sim2$}
\label{alg:simulate}
$M,N,S\leftarrow\emptyset$\;

\While {$\exists \, e^*=(\ell^*,r^*) = \argmax_{e\in E} \,\{ f_M(e) \,\mid \, M\cup
 \{e\} \mbox{ is a matching }\}$}
{
 Flip a coin with probability $p$ of head\;
 {\bf if} $head$ {\bf then} $M\leftarrow{}M\cup\{e\}$\; {\bf else} { $N\leftarrow{}N\cup\{e\}$}\;
Remove all edges incident to $\ell^*$ from $E$\;
}
\ForEach {edge $e=(\ell,r) \in N$}{
 Add $e$ to $S$ if $e$ is the only edge incident to $r$ in $N$\;
}
return $S$\;
\end{algorithm}

\subsection{Analysis}

We cast the stochastic process in $\sim2$ to our simple stochastic
process as follows. In particular, at each time $i$ an edge $e$ is
selected in $\sim2$, we define $w_i = f_{M}(e)$ where $M$ is the
current set of elements in $M$. Clearly, $w_i$ is a mapping from
previous $i-1$ Bernoulli random variables. Our process will terminate
after $n$ edges are selected. So $\mu = n$. In case the process
terminates before $|M|=n$, we can further append dummy edges in the
process.

Notice that $w(M) = f(M)$. By Proposition~\ref{pro:weight of M}, we have 
\begin{equation}
\label{eqn:transversal w(N)}
\E[w(N)] = \frac{1-p}{p}\E[w(M)] = \frac{1-p}{p} \E[f(M)].
\end{equation}


{\bf Pruning.} Since $N$ may not be a matching, we 
remove all edges in $N$ that share the same node in $R$ with other edges in $N$. Notice that no two edges in $N$ share the same left node.
 Let $S$ be the
set of edges left. Define $E_r$ be the set of edges incident to $r\in
R$. Then $|S\cap E_r|\leq 1$. We have
\[ N = S \cup \left( \cup_{r\in R} \one_{|N\cap E_r| >1} \cdot (N\cap
  E_r)\right),\]
and 
\begin{equation}
\label{eqn:all}
\E[w(N)] \leq \E[w(S)] + \sum_{r\in R} \E[ w(\one_{|N\cap E_r| >1} \cdot (N\cap
  E_r))].
\end{equation}

Now we focus on the term $\E[ w(\one_{|N\cap E_r| >1} \cdot (N\cap
  E_r))]$ for a particular node $r\in R$. We isolate our stochastic
  process on edges in $E_r$, by rearranging randomness as follows. For
  each edge in $e \in E\setminus E_r$, we associate a biased random
  coin. When $e$ arrives in the process, the random coin associated
  with it will be tossed. (Since all edges incident to the same left node will be processed only once on the first arriving edge, we will not toss two random coins for the same left node.)

For a fixed set of outcomes of random coins associated with edges in
$E\setminus E_r$, the process on edges in $E_r$ is a simple stochastic
process with $\mu =1$. Therefore, by Proposition~\ref{pro:weight of
  S}, we have
\begin{equation}
\label{eqn:singler}
\E[w(\one_{|N\cap E_r|>1} \cdot (N\cup E_r))] \leq \frac{1-p^2}{p^2}
\E[w(S\cap E_r)].
\end{equation}
Since $E_r$s are disjoint, and $w(\cdot)$ is linear for a fixed set of random
outcomes, we have
\begin{equation}
\label{eqn:sum}
\sum_{r\in R} w(S\cap E_r) = w(S)
\end{equation}

Combining Eqn.(\ref{eqn:all}), Eqn.(\ref{eqn:singler}) and
Eqn.(\ref{eqn:sum}), we immediately have:
\begin{equation}
\label{eqn:transversal w(S)}
\E[w(S)] \geq p^2\cdot\E[w(N)].
\end{equation}

Finally, we bound $f(S)$ based on $w(S)$ following an approach similar to the laminar matroid case. Again, we define $g(e) = f_{N_e}(e)$, i.e., if
$e$ appears in the greedy order, $N_e$ is the current set of elements
in $N$; otherwise, $g(e) = 0$.

\begin{lemma}
  \label{lem:transversal f(S)}
  For any $t>0$, let $\alpha=\frac{p}{1-p}$ and
  $\theta=\frac{t+\alpha}{\alpha}$, we have
  \[\E[f(S)]\geq(\frac{1}{\theta}-\frac{1}{t\cdot p^2})\cdot\E[w(S)]\]
\end{lemma}
\begin{proof}
Let $g(S)$ be the function $\sum_{e\in S}g(e)$. Since
$S\subset N$, we have $f(S)\geq g(S)$. We inspect the function
$F(S,M,N)=t\cdot g(S)+f(M)-f(M\cup N) \leq t\cdot g(S)$. 

Following exactly the same argument in the proof of
Lemma~\ref{lem:laminarS}, we have the same conclusion as in
Eqn.(\ref{eqn:savedforlater}):
\[t\cdot\E[f(S)]\geq\frac{t}{\theta}\E[w(S)]-\E[w(N)].\]

The lemma immediately follows from Eqn.(\ref{eqn:transversal w(S)}). 
\end{proof}
Combine Lemma~\ref{lem:transversal f(M)}, Eqn.(\ref{eqn:transversal w(N)}), Eqn.(\ref{eqn:transversal
  w(S)}) and Lemma~\ref{lem:transversal f(S)}, 
\[\E[f(S)]\geq(\frac{1}{\theta}-\frac{1}{t\cdot p^2})\cdot
p^2\cdot\frac{1-p}{3}\cdot\opt\geq \frac{1}{95}\opt\]
The inequality comes from taking $p=0.9$ and $t=5.29$. We have the main result of this section.

\begin{theorem}
\label{thm:transversal}
There is an online algorithm with competitive ratio at most 95 for the submodular matroid secretary problem with transversal matroids. 
\end{theorem}

\section{Conclusion}
\label{sec:conclusion}

In this paper, we develop a general algorithm for the submodular
matroid secretary problems. In particular, we obtain constant
competitive algorithms for laminar matroids and transversal
matroids. Our algorithm can also handle the intersection of a constant
number of laminar matroids, which makes our algorithm more
applicable. 

Our algorithm does not work on general matroids. Consider
the following simple example on graphical matroids. There is a single
heavy edge $(u,v)$ in the graph. There is a large number of nodes
$K=\{u_1,u_2,\ldots, u_n\}$ and edges $\{(u, u_i), (u_i, v) \mid u_i
\in K\}$. The weight on each such edge is very small. It is easy to
verify that the probability that our algorithm will accept $(u,v)$ is
exponentially small on $n$. Nevertheless, our algorithm can handle
graphical matroids using the same decomposition
technique~\cite{BabaioffDGIT09}, i.e., by reducing the problem to 
a partition matroid, which is randomly selected from two constructed partition matroids.
 On the other hand, it would be
interesting to characterize the independent set constraints for which
our algorithm framework is constant competitive.

In the distinction between the submodular case and linear case in
matroid secretary problem, we still cannot adapt the recent
$O(\sqrt{\log r})$ competitive algorithm in~\cite{ChakrabortyL2012} as well
as the constant competitive algorithm for the random assignment model
in~\cite{Soto11} previously on the linear case. It would be interesting to
close this gap. Finally, it is still widely open whether the matroid
secretary problem permits constant competitive algorithms for general
matroids.

\bibliographystyle{plain}
\bibliography{IS}

\newpage
\appendix
\input{appendix}
\end{document}

%% file: appendix.tex
\section{Missing Proofs in Section \ref{sec:pre}}
\label{app:pre}
\noindent{\textbf{Lemma \ref{lem:identical}} (restated).}{\emph\lemidentical}

\begin{proof}
We couple the randomness in $\sim$ and $\online$ as follows. In
$\online$, the following randomness is used: (a) the random
permutation $\pi_O$ of $U$; (b) a random number $k = Binom(|U|, p)
$. Let $H_O$ be the first $k$ elements in $\pi_O$. For a
permutation $\pi$ of $U$, let $H_\pi$ be a (non-ordered) prefix of $\pi$.
Then for any fixed permutation
$\pi$ and $H_\pi$,

\[ \Pr[ \pi_O = \pi \land H_O= H_\pi ] = (n!)^{-1}{n \choose |H_\pi|} p^{|H_\pi|}(1-p)^{n-|H_\pi|}.\]


In $\sim$, we can associate each element in $U$ with a biased coin
with head probability $p$. Let $H_S$ be the set of nodes in $U$ whose
coin is head and $T_S = U\setminus H_S$. 
We append the
randomness of $\sim$ by applying random permutations on $H_S$ and $T_S$. A
permutation of $U$ is the concatenation of $H_S$ and $T_S$ denoted as
$\pi_S$. 

\[ \Pr[\pi_S = \pi \land H_S = H_\pi] =  
p^{|H_\pi|}(1-p)^{n-|H_\pi|}( (|H_\pi|)!)^{-1} ( (n-|H_\pi|)!)^{-1}.
\]
 

Therefore, the probabilities of having a particular permutation and $H$
are the same in the two algorithms. It is then sufficient to show that
both algorithms generate the same $M$ and $N$, given a fixed
permutation $\pi$ of $U$ and $H$.

Notice that $M = \greedy(H)$ in both algorithms, which must be
identical. Now we prove for $N$. Let $N_O$ and $N_S$ be the $N$ in
$\online$ and $\sim$, respectively.

Consider element $e\in N_O$. By $\online$,  $e\in M' = \greedy(H\cup\{e\})$.
Assume $e$
is the $i$-th element added in $M'$. Let $M_{i-1}$ be the first $i-1$
elements placed into $M$ in $\online$. Notice that  the first
$i-1$ elements placed into $M$ in $\sim$ is exactly
$M_{i-1}$. Let $e_i$ be the
$i$-th element placed into $M$ in $\sim$.  Since $e\in M'$,
$f_{M_{i-1}}(e) > f_{M_{i-1}}(e_i)$. Therefore, in $\sim$, $e$ must be
processed before $e_i$ is placed into $M$. As the coin associated with
$e$ is tail, we conclude that $e\in N_S$.

For the other direction, 
consider $e\in N_S$. Let $M_e \subseteq M$ be the set of elements in $M$ when
$e$ is processed in $\sim$. By the greedy nature of $\sim$,
$f_{M_e}(e)$ is larger than any other elements in $H\setminus
\{M_e\}$. Therefore, $e\in \greedy(H\cup \{e\})$, i.e., $e\in N_O$.
\end{proof}

%
%
%
%
